\documentclass[conference,letterpaper]{IEEEtran}
\IEEEoverridecommandlockouts
\usepackage[T1]{fontenc}
\usepackage{cite}
\usepackage{amsmath,amssymb,amsfonts,amsthm}
\usepackage{graphicx,booktabs,tabularx,needspace}
\usepackage{balance}
\usepackage{tikz}
\usetikzlibrary{arrows.meta, decorations.pathreplacing,patterns}
\usepackage{xurl}
\usepackage[hidelinks]{hyperref}
\hypersetup{pdftitle={Priority-Aware Firm and Flexible Capacity Quantification for Large-Load Interconnection}}
\newtheorem{proposition}{Proposition}
\newtheorem{theorem}{Theorem}

\usepackage{xcolor}
\newcommand{\defeq}{\mathrel{:=}}
\newcommand{\R}{\mathbb R}
\newcommand{\E}{\mathbb E}
\newcommand{\Prob}{\mathbb P}
\newcommand{\one}{\mathbf 1}
\newcommand{\pos}[1]{[#1]_+}
\newcommand{\cvar}{\operatorname{CVaR}}
\newcommand{\T}{\mathcal T}
\newcommand{\I}{\mathcal I}
\newcommand{\B}{\mathcal N}
\newcommand{\Sc}{\mathcal S}
\newcommand{\J}{\mathcal J}
\newcommand{\K}{\mathcal K}
\newcommand{\M}{\mathcal M}
\title{Priority-Aware Firm and Flexible Capacity Quantification for Large-Load Interconnection}

\author{Shaoze Li$^{1}$, Bohang Fang$^{1}$,
Junjie Qin$^{2}$, and Cong Chen$^{1,*}$%
\thanks{\raggedright
$^{1}$Shaoze Li, Bohang Fang, and Cong Chen are with the
Thayer School of Engineering, Dartmouth College,
Hanover, NH, USA.
\texttt{\{Shaoze.Li.TH, Bohang.Fang, Cong.Chen\}@dartmouth.edu}. $^{2}$Junjie Qin is with the Elmore Family School of
Electrical and Computer Engineering, Purdue University,
West Lafayette, IN, USA.
\texttt{\{jq@purdue.edu\}}. \textit{*Corresponding author: Cong Chen.} 
}}

\begin{document}
\maketitle

\begin{abstract}

We study how to unlock existing transmission capacity to accelerate large-load interconnection under differentiated quality-of-service (QoS) requirements. We establish an efficient framework to quantify firm and flexible hosting capacities and assess marginal values of grid upgrades. Specifically, we define multiple flexible capacity tiers with differentiated QoS limits and design a two-layer priority rule to allocate curtailment among tiers. We then formulate a mixed-integer chance-constrained model to jointly quantify firm and flexible capacities under QoS and priority constraints. To improve computational efficiency, we derive an equivalent aggregate formulation that eliminates individual curtailment variables and priority binaries without changing the feasible capacity set.  Numerical results show the computational advantage of aggregation. Multi-tier flexible capacity increases total admitted capacity by 45.0\% over the firm-only hosting capacity.

\end{abstract}


\section{Introduction}\label{sec:intro}

The rapid growth of large electricity loads, including AI data centers, is placing increasing pressure on transmission grid interconnection~\cite{ferc2025colocation}. However, transmission grid expansion is slow and capital-intensive, making it difficult to keep pace with large-load interconnection requests.   This paper focuses on accommodating this demand surge before network upgrades while meeting differentiated quality-of-service (QoS) requirements. Specifically, we consider multi-tier flexible interconnection, where different tiers have distinct QoS limits. Unlike firm capacity, which must be served under all planning conditions, flexible capacity (also known as non-firm load service) allows curtailment subject to agreed QoS limits.




Existing studies demonstrate the benefits of such flexibility for load growth in U.S. power systems \cite{norris2025}, distribution-level interconnection \cite{gu2026role}, and data-center siting in transmission networks \cite{kim2026flexibility}. However, large-load customers may differ in their QoS requirements, both in terms of allowed curtailment depth and the probability of exceeding that depth. These studies do not explicitly consider multiple flexible capacity tiers with differentiated QoS requirements.

Another challenge is how to allocate curtailment when multiple flexible tiers share a bus. Their QoS limits do not specify how curtailment should be distributed among them. Priority-based allocation provides one efficient approach and has been studied in electricity markets \cite{chao1987priority}. PJM has also proposed curtailment priorities for certain large-load services \cite{pjm2026iras}. However, these studies do not determine how much firm and tier-specific flexible capacity a transmission network can jointly admit while satisfying network and QoS constraints.

Our contributions addressing these gaps are twofold. First, we develop a two-layer curtailment priority rule and formulate a unified mixed-integer chance-constrained model that maximizes the preference-weighted firm and flexible interconnection capacity subject to network, priority, and QoS limits. We then establish an exact aggregate reformulation that eliminates individual curtailment variables and priority binaries without changing the feasible set of capacity decisions. 
Second, to improve computational efficiency and guide grid upgrades, we first approximate the aggregate model by a linear program (LP) whose feasible solutions provably satisfy the QoS limits, then derive marginal upgrade values from the dual variables that indicate how much additional hosting capacity can be unlocked by the upgrade in transmission or generation capacity. Numerical results on the PJM 5-bus system show that aggregation reduces solution time from 5801.5 s to 1.56 s while preserving the optimal objective value. On the IEEE RTS 73-bus system, results show that multi-tier flexible capacity increases total admitted capacity by 45.0\% over the firm-only hosting capacity. 

The paper is organized as follows. Section~\ref{sec:individual} defines the capacity products and formulates the individual model under the two-layer curtailment priority rule.
Section~\ref{sec:projection} derives the equivalent aggregate model and the LP approximation, whose duals give marginal upgrade values.
The experiments are given in Section~\ref{sec:numerics}.

\section{Individual  Capacity Quantification}\label{sec:individual}

In this section, we first introduce the QoS limits and the curtailment priority rule, and then formulate an individual model for capacity quantification. The grid planner performs this quantification to determine firm and flexible interconnection capacities.  In the next section, we derive an equivalent aggregate model and the LP approximation to improve computational efficiency and obtain marginal upgrade information.
\subsection{Firm and Flexible Capacity Products}
Let $\B$ and $\J$ denote the sets of buses and transmission lines, respectively, and let $\I\subseteq\B$ denote the candidate interconnection buses. The planning horizon is $\T\defeq \{1,\ldots,T\}$, and $\K\defeq \{1,\ldots,K\}$ indexes flexible service tiers with different QoS requirements. Existing nodal demand is represented by a random trajectory $\mathbf D\defeq (\mathbf D_t)_{t\in\T}$, where $\mathbf D_t\defeq(D_{it})_{i\in\B}\in\R^{|\B|}$. Throughout the paper, $\Prob$ and $\E$ denote probability and expectation under the distribution of $\mathbf D$. Unless otherwise specified, $i\in\I$, $k\in\K$, and $t\in\T$ throughout the formulations. Bold symbols denote vectors and matrices, and vector inequalities are componentwise. Unindexed bold decision symbols collect the corresponding variables over all relevant indices.

At bus $i$, \emph{firm capacity} $c_i^{\mathrm{F}}\ge0$ is served in full in every period and almost every realization. \emph{Flexible capacity} $c_{ik}\ge0$ belongs to service tier $k$ and may be curtailed under specified conditions. Let $\mathbf c_i\defeq(c_{ik})_{k\in\K}$ and $\mathbf c_k\defeq(c_{ik})_{i\in\B}$ collect flexible capacities across tiers and buses, respectively, and let $\mathbf c\defeq(c_{ik})_{i\in\I,k\in\K}$ collect all flexible capacity awards. The firm-capacity vector is $\mathbf c^{\mathrm{F}}\defeq(c_i^{\mathrm{F}})_{i\in\B}$. Each admitted flexible product requests its full capacity in every period. Product $(i,k)$ refers to service tier $k$ at bus $i$. Let $y_{ikt}$ denote the curtailed amount of product $(i,k)$ in period $t$. The delivered flexible load is then $c_{ik}-y_{ikt}$, where $0\le y_{ikt}\le c_{ik}$. Define $\mathbf y_{kt}\defeq(y_{ikt})_{i\in\B}$ for the tier-$k$ curtailment vector.
New-load capacities, request limits, and associated curtailments are extended by zero to buses outside $\I$. 

Each QoS tier $k$ is defined by two parameters $(\delta_{k},\epsilon_{k})$. The parameter $\delta_{k}$ determines the allowed curtailment depth $\delta_{k}c_{ik}$, while $\epsilon_{k}$ limits the probability that this depth is exceeded at least once during the planning horizon:
\vspace{-4.5pt}
\begin{equation}
\Prob\!\left(\max_{t\in\T}y_{ikt}>\delta_{k}c_{ik}\right)
\le\epsilon_{k}.\label{eq:contract}
\end{equation}
\par
\vspace{-5pt}
\noindent
 Curtailment within the allowed depth is not frequency-limited. These QoS parameters are specified in the interconnection agreement, allowing customers to plan on-site backup resources or load flexibility based on the agreed curtailment depth, while limiting the probability of deeper curtailment. Together, these provisions enable earlier interconnection under differentiated QoS requirements.
 
 The tiers are ordered from most to least reliable, with
\vspace{-4.5pt}
\begin{equation}
\begin{split}
0\le\delta_1\le\cdots\le\delta_K<1,\\
0<\epsilon_1\le\cdots\le\epsilon_K<1.
\end{split}\label{eq:order}
\end{equation}
\par
\vspace{-5pt}
\noindent
Thus, a more reliable tier has no larger allowed curtailment depth or exceedance-probability limit. We use common parameters $(\delta_{k},\epsilon_{k})$ across buses for notational simplicity; the results also extend to bus-specific reliability parameters satisfying~\eqref{eq:order} at each bus. A flexible product $(i,k)$ is called active if $c_{ik}>0$ and inactive otherwise.

\subsection{The Two-Layer Curtailment Priority Rule}\label{subsec:priority}
  
\begin{figure}[!ht]
\centering
\begin{tikzpicture}[x=0.32cm,y=0.40cm,scale=1,transform shape,>=Stealth,font=\scriptsize]
\draw[->] (0,0) -- (21.2,0);
\node[below] at (10.6,-2.5) {Total curtailment (MW)};
\draw[->] (0,0) -- (0,5.45);
\node[rotate=90,anchor=south] at (-1.05,2.8) {Curtailment step};
\fill[gray!14] (0,0.85) rectangle (5,1.15);
\fill[gray!14] (5,1.85) rectangle (7,2.15);
\fill[pattern=north east lines,pattern color=gray] (7,2.85) rectangle (12,3.15);
\fill[pattern=north east lines,pattern color=gray] (12,3.85) rectangle (20,4.15);
\draw[very thick] (0,1)--(5,1)--(5,2)--(7,2)--(7,3)--(12,3)--(12,4)--(20,4);
\foreach \x/\lab in {0/0,5/5,7/7,12/12,20/20}{\draw (\x,0.1)--(\x,-0.1) node[below]{\lab};}
\foreach \y in {1,2,3,4}{\draw (0,\y)--(-0.16,\y) node[left]{\y};}
\node[above,align=center] at (2.5,1.15) {$a_2=5$\\Product 2};
\node[above,align=center] (aone) at (5.1,3.65) {$a_1=2$\\Product 1};
\draw[thin,->] (aone.south) -- (6,2.08);
\node[above,align=center] at (9.5,3.15) {$b_2=5$\\Product 2};
\node[above,align=center] at (16,4.15) {$b_1=8$\\Product 1};
\draw[decorate,decoration={brace,mirror,amplitude=3pt}] (0,-1.3)--(7,-1.3);
\node[below] at (3.5,-1.5) {Within-depth layer};
\draw[decorate,decoration={brace,mirror,amplitude=3pt}] (7,-1.3)--(20,-1.3);
\node[below] at (13.5,-1.5) {Exceedance layer};
\end{tikzpicture}
\vspace{-6pt}
\caption{A toy example for two-layer curtailment priority with $c_1=c_2=10$~MW, $\delta_1=0.2$, and $\delta_2=0.5$, with the bus index omitted. Product~2 is less reliable. Curtailment follows the order $a_2\rightarrow a_1\rightarrow b_2\rightarrow b_1$. Tranche widths give the amount curtailed in each step. }
\label{fig:stack}
\vspace{-6pt}
\end{figure}
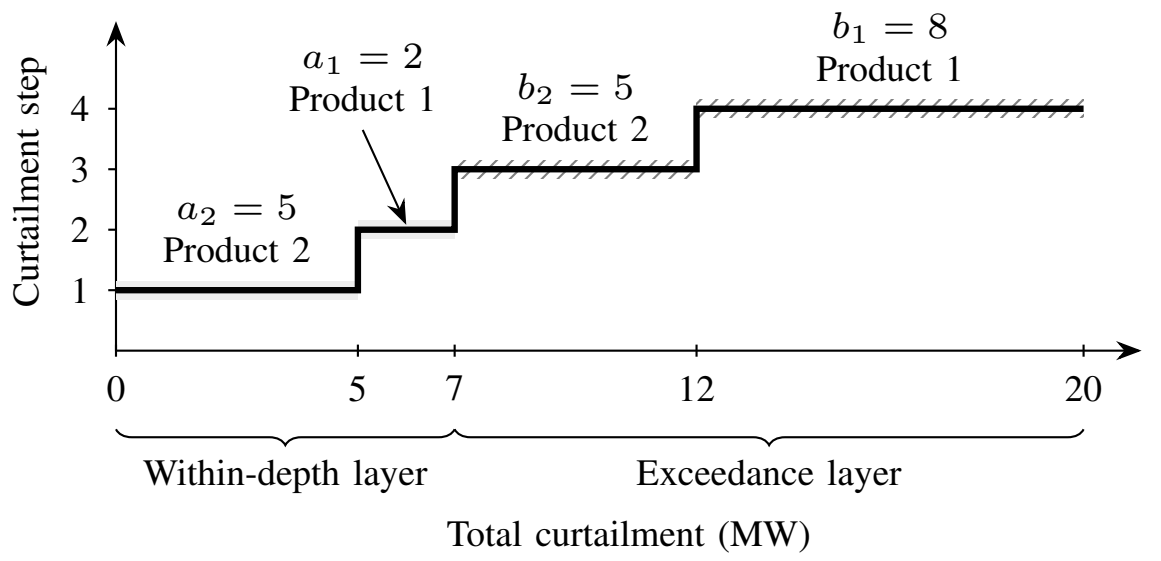

Equation~\eqref{eq:contract} specifies the QoS requirement of each flexible product but does not determine how curtailment is allocated. We therefore adopt a priority rule for products sharing a bus. Specifically, we adopt a two-layer rule: the allowed curtailment depths of all flexible products are used before any product is curtailed beyond its allowed depth. The first layer is the within-depth layer, with tranche sizes $a_{ik}\defeq\delta_{k}c_{ik}$; the second layer is the exceedance layer, with tranche sizes $b_{ik}\defeq(1-\delta_{k})c_{ik}$. Within each layer, less reliable tiers are curtailed before more reliable ones. 
Next, a small example of two products shows the curtailment priority. Fig.~\ref{fig:stack} illustrates this order for two products, with the bus index omitted. Curtailment first uses the 5~MW within-depth tranche of product~2, then the 2~MW within-depth tranche of product~1. Only after both within-depth tranches are exhausted does it use the 5~MW exceedance tranche of product~2, followed by the 8~MW of product~1.
\vspace{-0.8em}

Let $\M\defeq \{1,\ldots,2K\}$ index all curtailment tranches at a bus. Following the two-layer rule, these tranches are curtailed in the order
\vspace{-4.5pt}
\begin{equation}
(d_{i1}(\mathbf c_i),\ldots,d_{i,2K}(\mathbf c_i))
\defeq (a_{iK},\ldots,a_{i1},b_{iK},\ldots,b_{i1}).\label{eq:stack}
\end{equation}
\par
\vspace{-5pt}
\noindent
Here, $d_{im}(\mathbf c_i)$ denotes the size of tranche $m$, which can be expressed as
\vspace{-4.5pt}
\[
d_{im}(\mathbf c_i)\defeq
\begin{cases}
\delta_{K-m+1}c_{i,K-m+1},&1\le m\le K,\\[2pt]
(1-\delta_{2K-m+1})c_{i,2K-m+1},&K<m\le2K.
\end{cases}
\]
\par
\vspace{-5pt}
\noindent
Each tranche size is linear in the flexible capacity decisions. Under this ordering, tranche $m+1$ can receive curtailment only after tranche $m$ has been fully curtailed.

We next formulate the two-layer priority rule using linear constraints and binary variables. Let $x_{imt}$ denote the curtailed amount assigned to tranche $m$ in period $t$. To enforce the curtailment order, we introduce binary variables $\beta_{imt}$ for $m=0,\ldots,2K$, with fixed endpoints $\beta_{i0t}=1$ and $\beta_{i,2K,t}=0$. For $0<m<2K$, setting $\beta_{imt}=1$ indicates that tranche $m$ is fully curtailed, allowing the next tranche to receive curtailment. The priority rule is then represented by the following constraints:
\vspace{-4.5pt}
\begin{subequations}\label{eq:priority-representation}
\begin{alignat}{2}
&0\le x_{imt}\le d_{im}(\mathbf c_i),
&&\quad\forall i,m,t,\label{eq:tranche-bounds}\\
&x_{imt}\le M\beta_{i,m-1,t},
&&\quad\forall i,m,t,\label{eq:tranche-open}\\
&x_{imt}\ge d_{im}(\mathbf c_i)-M(1-\beta_{imt}),
&&\quad\forall i,m,t,\label{eq:tranche-full}\\
&\beta_{i,m-1,t}\ge\beta_{imt},
&&\quad\forall i,m,t,\label{eq:binary-chain}\\
&y_{ikt}=x_{i,K-k+1,t}+x_{i,2K-k+1,t},
&&\quad\forall i,k,t,\label{eq:product-recovery-row}\\
&\beta_{i,0,t}=1,\quad\beta_{i,2K,t}=0,
&&\quad\forall i,t,\label{eq:binary-endpoints}\\
&\beta_{imt}\in\{0,1\},
&&\quad\forall i,m,t.\label{eq:priority-domains}
\end{alignat}
\end{subequations}
\par
\vspace{-5pt}

Let $\bar C_i^{\mathrm{F}}$ and $\bar C_{ik}$ denote the maximum firm and tier-$k$ capacity requests at bus $i$, respectively, and set $\bar{\mathbf c}_i\defeq(\bar C_{i1},\ldots,\bar C_{iK})$. A valid big-M constant is $M\defeq\max_{i,m}d_{im}(\bar{\mathbf c}_i)$, which bounds every feasible tranche size. Constraints~\eqref{eq:tranche-bounds}--\eqref{eq:binary-chain}, together with the fixed endpoints, ensure that a tranche receives curtailment only after all earlier tranches are fully curtailed, leaving at most one partially curtailed tranche. Equation~\eqref{eq:product-recovery-row} recovers each product's curtailment by summing its two tranche curtailments, ensuring $0\le y_{ikt}\le c_{ik}$.

\subsection{Mixed-Integer Chance-Constrained Quantification Model}

Combining the curtailment priority rule with the generation and transmission network constraints, we formulate the following model to determine the firm and flexible capacities. The capacities $\mathbf c^{\mathrm{F}}$ and $\mathbf c$ are first-stage decisions made before $\mathbf D$ is realized; $\mathbf g$, $\mathbf q$, $\mathbf y$, $\mathbf x$, and $\boldsymbol\beta$ are recourse decision rules depending on $\mathbf D$. We suppress this dependence and require the operational and priority constraints to hold almost surely.
\vspace{-2.5pt}
\begin{subequations}\label{eq:individual-problem}
\begin{IEEEeqnarray}{r'l'r}
\IEEEeqnarraymulticol{3}{l}{
(\mathcal{P}_{\mathrm{ind}})\;
\max_{\substack{
\mathbf c^{\mathrm{F}},\mathbf c,\mathbf g,\mathbf q,\\
\mathbf y,\mathbf x,\boldsymbol\beta}}
W(\mathbf c^{\mathrm{F}},\mathbf c)
:= (\mathbf w^{\mathrm{F}})^\top\mathbf c^{\mathrm{F}}
+\sum_{k\in\K}\mathbf w_k^\top\mathbf c_k,
\;\IEEEeqnarraynumspace
}\label{eq:individual-objective}\\
\text{s.t.}
&\mathbf 0\le\mathbf c^{\mathrm{F}}\le\bar{\mathbf C}^{\mathrm{F}},\;
 \mathbf 0\le\mathbf c_k\le\bar{\mathbf C}_k,
&\forall k,\quad\IEEEeqnarraynumspace\label{eq:bounds}\\
&\mathbf q_t=\mathbf g_t-\mathbf D_t-\mathbf c^{\mathrm{F}}
&\IEEEnonumber\\*
&\quad-\sum_{k\in\K}(\mathbf c_k-\mathbf y_{kt}),
&\forall t,\quad\IEEEeqnarraynumspace\label{eq:individual-injection}\\
&\one^\top\mathbf q_t=0,
&\forall t,\quad\IEEEeqnarraynumspace\label{eq:balance}\\
&-\bar{\mathbf f}\le\mathbf S\mathbf q_t\le\bar{\mathbf f},
&\forall t,\quad\IEEEeqnarraynumspace\label{eq:line}\\
&\underline{\mathbf g}\le\mathbf g_t\le\bar{\mathbf g},
&\forall t,\quad\IEEEeqnarraynumspace\label{eq:generation}\\
&-\boldsymbol\Delta^-\le\mathbf g_t-\mathbf g_{t-1}\le\boldsymbol\Delta^+,
&\forall t,\quad\IEEEeqnarraynumspace\label{eq:ramp}\\
&\Prob\!\left(\max_{t\in\T}y_{ikt}>\delta_{k}c_{ik}\right)\le\epsilon_{k},
&\forall i,k,\quad\IEEEeqnarraynumspace\label{eq:individual-risk-row}\\
&\text{Priority constraints \eqref{eq:priority-representation}},
&\IEEEeqnarraynumspace\label{eq:priority-model}
\end{IEEEeqnarray}
\end{subequations}
\par
\vspace{-5pt}
\noindent
where $\mathbf w^{\mathrm{F}}\defeq(w_i^{\mathrm{F}})_{i\in\B}$ and $\mathbf w_k\defeq(w_{ik})_{i\in\B}$ collect the preference weights for the capacities, and $\mathbf g_t\defeq(g_{it})_{i\in\B}$ and $\mathbf q_t\defeq(q_{it})_{i\in\B}$ denote nodal generation and net injection.  The objective~\eqref{eq:individual-objective} maximizes the  preference-weighted admitted capacity. Unit weights give the total admitted capacity in MW. Constraint~\eqref{eq:bounds} imposes the capacity request limits. Constraints~\eqref{eq:individual-injection}--\eqref{eq:ramp} describe the lossless DC network model. Constraint~\eqref{eq:individual-injection} defines net injection, and~\eqref{eq:balance} enforces power balance, where $\one$ is the all-ones vector. Constraint~\eqref{eq:line} limits transmission flows, where $\mathbf S\in\R^{|\J|\times|\B|}$ is the power transfer distribution factor (PTDF) matrix and $\bar{\mathbf f}$ denotes the line ratings. Constraints~\eqref{eq:generation} and~\eqref{eq:ramp} impose the generation limits $\underline{\mathbf g}$ and $\bar{\mathbf g}$ and the ramp-down/up limits $\boldsymbol\Delta^-$ and $\boldsymbol\Delta^+$, respectively, with $\mathbf g_0$ denoting the given initial dispatch.  Constraint~\eqref{eq:individual-risk-row} imposes product-level QoS limits defined in~\eqref{eq:contract}, while~\eqref{eq:priority-model} enforces the curtailment priority. 



\section{Aggregate Reformulation}\label{sec:projection}

Model $\mathcal{P}_{\mathrm{ind}}$ captures the curtailment priority and the QoS requirements of each flexible product. However, the priority binaries and chance constraints make the model computationally challenging.  The key observation is that the network constraints~\eqref{eq:individual-injection}--\eqref{eq:ramp} depend on the flexible products only through their total capacity and total curtailment at each bus. Inspired by this observation, we first develop an aggregate model $\mathcal{P}_{\mathrm{agg}}$ and prove that it is equivalent to $\mathcal{P}_{\mathrm{ind}}$. This can improve computational efficiency, with a simulation comparison in Table~\ref{tab:pjm}. We then apply a conservative conditional value-at-risk (CVaR) approximation to the remaining chance constraints to obtain a linear program (LP). The dual information of this LP can then be used to identify candidate upgrades.

We begin by defining the total flexible capacity and aggregate curtailment at each bus $i\in\B$:
\vspace{-4.5pt}
\begin{equation}
C_i(\mathbf c_i)\defeq \sum_{k\in\K}c_{ik},\qquad
z_{it}\defeq \sum_{k\in\K}y_{ikt}.\label{eq:aggregate-def}
\end{equation}
\par
\vspace{-5pt}
\noindent
Let $\mathbf C(\mathbf c)\defeq (C_i(\mathbf c_i))_{i\in\B}$ and $\mathbf z_t\defeq (z_{it})_{i\in\B}$ denote the corresponding vectors. Using~\eqref{eq:aggregate-def} in~\eqref{eq:individual-injection}, together with $0\le y_{ikt}\le c_{ik}$, gives
\vspace{-4.5pt}
\begin{subequations}\label{eq:aggregate-physics}
\begin{align}
&\mathbf q_t=\mathbf g_t-\mathbf D_t-\mathbf c^{\mathrm{F}}
-\mathbf C(\mathbf c)+\mathbf z_t,\quad\forall t,\label{eq:aggregate-injection}\\
&\mathbf 0\le\mathbf z_t\le\mathbf C(\mathbf c),\quad\forall t.\label{eq:z-bounds}
\end{align}
\end{subequations}
\par
\vspace{-5pt}
\noindent
In the aggregate model, $\mathbf z_t$ is a recourse decision vector.

\subsection{Aggregate Representation of Curtailment Priority}

To preserve the curtailment priority in the aggregate model, we first show that the curtailment of each product can be recovered from the total curtailment at its bus. Define
\vspace{-4.5pt}
\begin{equation}
\Gamma_{im}(\mathbf c_i)\defeq\sum_{h=1}^{m}d_{ih}(\mathbf c_i),\qquad m=0,\ldots,2K,\label{eq:cumulative}
\end{equation}
\par
\vspace{-5pt}
\noindent
as the cumulative size of the first $m$ tranches. Following~\eqref{eq:stack}, $m$ ranges from 0 to $2K$ because the two-layer priority rule has two tranches for each of the $K$ tiers. Thus, tranche $m$ occupies the interval $[\Gamma_{i,m-1},\Gamma_{im}]$ of aggregate curtailment, where $\Gamma_{i0}=0$ and $\Gamma_{i,2K}=C_i$. From \eqref{eq:stack}, the within-depth tranche $a_{ik}$ of product $k$ is at position $K-k+1$, while its exceedance tranche $b_{ik}$ is at position $2K-k+1$. Their starting points are
\vspace{-4.5pt}
\begin{equation}
\begin{split}
A_{ik}(\mathbf c_i)&\defeq\Gamma_{i,K-k}(\mathbf c_i)=\sum_{r>k}\delta_{r}c_{ir},\\
H_{ik}(\mathbf c_i)&\defeq\Gamma_{i,2K-k}(\mathbf c_i)=\sum_{r\le k}\delta_{r}c_{ir}+\sum_{r>k}c_{ir}.
\end{split}\label{eq:threshold}
\end{equation}
\vspace{-5pt}

Both $A_{ik}$ and $H_{ik}$ are linear in $\mathbf c_i$. For the example in Fig.~\ref{fig:stack}, $H_2=a_2+a_1=7$~MW and $H_1=a_2+a_1+b_2=12$~MW. Thus, product~2 exceeds its allowed depth when total curtailment exceeds 7~MW, and product~1 exceeds its allowed depth when total curtailment exceeds 12~MW.

The following proposition shows that
aggregate curtailment uniquely determines individual
curtailments under the two-layer priority rule. We fix a bus $i$, a period, and a realization, and omit the period index and the argument $\mathbf c_i$ of $d_{im}$, $\Gamma_{im}$, $A_{ik}$, and $H_{ik}$. The positive part is defined as $\pos{u}:=\max\{u,0\}$.


\begin{proposition}[Recovery of individual curtailment]
\label{prop:stack}
Let the capacities satisfy~\eqref{eq:bounds}.
For every $z\in[0,C_i]$, there exists an allocation
satisfying~\eqref{eq:priority-representation} and
$\sum_{k} y_{ik}=z$.
All such allocations have the same tranche and product
curtailments, given by
\vspace{-4.5pt}
\begin{align}
x_{im}
&:=\min\big\{\pos{z-\Gamma_{i,m-1}},\,d_{im}\big\},
\quad m\in\M,
\label{eq:tranche-recovery}\\
y_{ik}&:=f_{ik}(z;\mathbf c_i)
:=\min\big\{\pos{z-A_{ik}},\,a_{ik}\big\}
\nonumber\\
&\quad+\min\big\{\pos{z-H_{ik}},\,b_{ik}\big\}.
\label{eq:recovery}
\end{align}
\par
\vspace{-5pt}
\end{proposition}
The proof is given in Appendix~\ref{app:proof-stack}.

We next show that the chance constraint of each product can also be expressed using aggregate curtailment.  Let $Z_i\defeq \max_{t\in\T}z_{it}$ denote the maximum aggregate curtailment at bus $i$ over the planning horizon.


{\begin{proposition}[Equivalent chance constraints]\label{prop:event}
Let $(\mathbf y,\mathbf x,\boldsymbol\beta)$ satisfy~\eqref{eq:priority-representation} in every realization, and let $z_{it}\defeq\sum_{k}y_{ikt}$. Then, for every active product $(i,k)$ with $c_{ik}>0$ and every realization,
\vspace{-4.5pt}
\begin{equation}
\max_{t\in\T}y_{ikt}>\delta_{k}c_{ik}
\quad\Longleftrightarrow\quad
Z_i>H_{ik}(\mathbf c_i).\label{eq:event}
\end{equation}

\noindent
Consequently, $\Prob(\max_{t\in\T}y_{ikt}>\delta_{k}c_{ik})=\Prob(Z_i>H_{ik}(\mathbf c_i))$, and the individual chance constraint~\eqref{eq:individual-risk-row} of product $(i,k)$ holds if and only if
$\Prob\big(Z_i>H_{ik}(\mathbf c_i)\big)\le\epsilon_{k}$.

\end{proposition}}
The proof is given in Appendix~\ref{app:proof-event}.

Proposition~\ref{prop:event} is the key step in the feasibility argument of Theorem~\ref{thm:equivalence} below: it allows the product-level chance constraints to be imposed directly on the aggregate curtailment.

 Next, we replace the individual curtailment variables with aggregate curtailment variables as follows:
\vspace{-2.5pt}
\begin{subequations}\label{eq:aggregate-problem}
\begin{align}
(\mathcal{P}_{\mathrm{agg}})\ 
&\max_{\mathbf c^{\mathrm{F}},\mathbf c,\mathbf g,\mathbf q,\mathbf z}\ W(\mathbf c^{\mathrm{F}},\mathbf c)\label{eq:aggregate-objective}\\
\text{s.t.}\ 
&\eqref{eq:bounds},\ \eqref{eq:balance}\text{--}\eqref{eq:ramp},\ \eqref{eq:aggregate-physics},\label{eq:aggregate-constraints}\\
&\Prob\big(Z_i>H_{ik}(\mathbf c_i)\big)\le\epsilon_{k},\quad\forall i,k.\label{eq:aggregate-risk}
\end{align}
\end{subequations}
\par
\vspace{-5pt}

\begin{theorem}
\label{thm:equivalence}
Under~\eqref{eq:order}, $\mathcal{P}_{\mathrm{ind}}$ and
$\mathcal{P}_{\mathrm{agg}}$ have the same feasible set of capacity decisions
$(\mathbf c^{\mathrm{F}},\mathbf c)$ and 
the same optimal value.
\end{theorem}
The proof is given in Appendix~\ref{app:proof-equivalence}.

The intuition behind Theorem~\ref{thm:equivalence} is that the network constraints depend on the flexible products only through
their total capacity and curtailment at each bus. For fixed capacities,
the priority rule uniquely determines the individual curtailments
from aggregate curtailment, and each active product's depth-exceedance
event is exactly captured by the aggregate threshold condition
in~\eqref{eq:aggregate-risk}. These properties eliminate the individual curtailment variables and priority binaries without changing the optimal value.

\subsection{CVaR-Based Linear Program}\label{sec:lp}

Although the aggregate model eliminates the priority binaries, its chance constraints~\eqref{eq:aggregate-risk} remain nonconvex. For each product $(i,k)$, define the loss $L_{ik}\defeq Z_i-H_{ik}(\mathbf c_i)$, so that~\eqref{eq:aggregate-risk} becomes $\Prob(L_{ik}>0)\le\epsilon_k$. We conservatively replace this chance constraint with $\cvar_{1-\epsilon_k}(L_{ik})\le0$, where $\cvar_{1-\epsilon}(L)$ denotes the upper-tail conditional value-at-risk (CVaR) at confidence level $1-\epsilon$. 

To obtain a finite LP, we consider $S$ demand trajectories $\mathbf D^s\defeq(\mathbf D_t^s)_{t\in\T}$, indexed by $s\in\Sc\defeq\{1,\ldots,S\}$, with probabilities $p_s>0$ and $\sum_{s\in\Sc}p_s=1$. The superscript $s$ also indexes the corresponding recourse variables, with $\mathbf g_0^s\defeq\mathbf g_0$. Applying the scenario-based CVaR reformulation~\cite{chen2024dera} yields the following LP.



\vspace{-5.5pt}
\begin{subequations}\label{eq:capacity-lp}
\interdisplaylinepenalty=0\relax
\begin{IEEEeqnarray}{l'r}
\IEEEeqnarraymulticol{2}{l}{
(\mathcal{P}_{\mathrm{LP}})\quad
\max_{\substack{
\mathbf c^{\mathrm{F}},\mathbf c,\mathbf g,\mathbf q,\\
\mathbf z,\boldsymbol\tau,\boldsymbol\xi}}
W(\mathbf c^{\mathrm{F}},\mathbf c)}\label{eq:lp-objective}\\
\text{s.t.}\quad \mathbf q_t^s=\mathbf g_t^s-\mathbf D_t^s-\mathbf c^{\mathrm{F}}
&\IEEEnonumber\\
\quad\quad-\mathbf C(\mathbf c)+\mathbf z_t^s,
&\forall t,s,\IEEEeqnarraynumspace\label{eq:lp-injection}\\
\quad\quad\mathbf 0\le\mathbf z_t^s\le\mathbf C(\mathbf c),
&\forall t,s,\IEEEeqnarraynumspace\label{eq:lp-z}\\
\quad\quad\one^\top\mathbf q_t^s=0,
&\forall t,s,\IEEEeqnarraynumspace\label{eq:lp-balance}\\
\quad\quad-\bar{\mathbf f}\le\mathbf S\mathbf q_t^s\le\bar{\mathbf f},
&\forall t,s,\IEEEeqnarraynumspace\label{eq:lp-line}\\
\quad\quad\underline{\mathbf g}\le\mathbf g_t^s\le\bar{\mathbf g},
&\forall t,s,\IEEEeqnarraynumspace\label{eq:lp-generation}\\
\quad\quad-\boldsymbol\Delta^-\le\mathbf g_t^s-\mathbf g_{t-1}^s\le\boldsymbol\Delta^+,
&\forall t,s,\IEEEeqnarraynumspace\label{eq:lp-ramp}\\
\quad\quad\xi_{ik}^s\ge z_{it}^s-H_{ik}(\mathbf c_i)-\tau_{ik},
&\forall i,k,t,s,\IEEEeqnarraynumspace\label{eq:epigraph1}\\
\quad\quad\tau_{ik}+\frac{1}{\epsilon_{k}}\sum_{s\in\Sc}p_s\xi_{ik}^s\le0,
&\forall i,k,\IEEEeqnarraynumspace\label{eq:epigraph2}\\
\quad\quad\xi_{ik}^s\ge0,\quad\eqref{eq:bounds},
&\forall i,k,s.\IEEEeqnarraynumspace\label{eq:excess-domains}
\end{IEEEeqnarray}
\end{subequations}
\par
\vspace{-4pt}
\noindent
To express the CVaR constraints in linear form, we introduce $\tau_{ik}\in\R$ and $\xi_{ik}^s\ge0$, collected in $\boldsymbol\tau$ and $\boldsymbol\xi$. The exact linearization of the finite-scenario CVaR formulation is derived in Appendix~\ref{app:cvar-lp-derivation}.

\begin{proposition}\label{thm:cvar-guarantee}
For the same finite-scenario distribution, let $(\mathbf c^{\mathrm{F}},\mathbf c,\mathbf g,\mathbf q,\mathbf z,\boldsymbol\tau,\boldsymbol\xi)$ be feasible for $\mathcal{P}_{\mathrm{LP}}$. Then $(\mathbf c^{\mathrm{F}},\mathbf c,\mathbf g,\mathbf q,\mathbf z)$ is feasible for $\mathcal{P}_{\mathrm{agg}}$. Consequently, the individual curtailments recovered by Proposition~\ref{prop:stack} form a feasible solution of $\mathcal{P}_{\mathrm{ind}}$ and satisfy~\eqref{eq:contract} for every product.
\end{proposition}
The proof is given in Appendix~\ref{app:proof-cvar}.

Model $\mathcal{P}_{\mathrm{LP}}$ is tractable, and its feasible solutions satisfy the sampled product-level QoS limits by Proposition~\ref{thm:cvar-guarantee}.
The vector pairs $(\underline{\boldsymbol\nu}_t^{s,*},\bar{\boldsymbol\nu}_t^{s,*})$, $(\underline{\boldsymbol\phi}_t^{s,*},\bar{\boldsymbol\phi}_t^{s,*})$, and $(\underline{\boldsymbol\psi}_t^{s,*},\bar{\boldsymbol\psi}_t^{s,*})$ denote the nonnegative optimal dual multipliers for the lower and upper line \eqref{eq:lp-line}, generation \eqref{eq:lp-generation}, and ramp bounds \eqref{eq:lp-ramp}.
By the envelope theorem, the marginal values of relaxing both bounds for line $\ell\in\J$, generation, and ramping at bus $n\in\B$ are
$\lambda_\ell\defeq\sum_{t,s}(\underline\nu_{\ell t}^{s,*}+\bar\nu_{\ell t}^{s,*})$,
$\lambda_n^{\mathrm G}\defeq\sum_{t,s}(\underline\phi_{nt}^{s,*}+\bar\phi_{nt}^{s,*})$, and
$\lambda_n^{\mathrm R}\defeq\sum_{t,s}(\underline\psi_{nt}^{s,*}+\bar\psi_{nt}^{s,*})$.


\section{Numerical Study}\label{sec:numerics}

We use the PJM 5-bus system to examine model equivalence, sensitivity to QoS parameters, and upgrade values, and the IEEE RTS 73-bus system to assess scalability.

\subsection{PJM 5-Bus System}

We first consider the PJM 5-bus system~\cite{li2010small}, with a data center at bus D. The five PJM buses are denoted by A, B, $\ldots$, E. Nominal bus loads are independently scaled by $1+\eta$, where $\eta\sim\mathcal N(0,0.05^2)$ is truncated to $[-0.10,0.10]$. For the three QoS tiers, the data center requests up to 200~MW of firm capacity and 120~MW per flexible tier. We set $\boldsymbol\delta=(0.01,0.05,0.10)$, $\boldsymbol\epsilon=(0.05,0.10,0.15)$, $w^{\mathrm{F}}=1$, and $\mathbf w=(0.9,0.8,0.7)$. These weights express relative preferences for capacity at different reliability levels. All models use the empirical distribution with $p_s=1/S$, and models in each comparison use the same demand samples. We use $S=500$ for Table~\ref{tab:pjm} and $S=2000$ for the other experiments. We ran all experiments using Gurobi 13.0.3 (default settings) on a Windows PC with a 20-core Intel Core Ultra 7 265 processor and 32 GiB of RAM.


To assess the benefit of aggregation, we compare $\mathcal{P}_{\mathrm{ind}}$ and $\mathcal{P}_{\mathrm{agg}}$. Table~\ref{tab:pjm} reports the same globally optimal objective value $W^\star$ for both models, as predicted by Theorem~\ref{thm:equivalence}; their capacity awards also coincide in this test. Aggregation removes the 2500 priority binaries. The remaining 1500 binaries enforce the chance constraints. 

Then, we use the duals of $\mathcal{P}_{\mathrm{LP}}$ to evaluate PJM upgrades ($S=2000$). We compute the marginal upgrade values in Table~\ref{tab:duals}. Ramp limits are nonbinding in this case. Table~\ref{tab:duals} identifies line D--E as the most valuable marginal upgrade, followed by generation at D, C, and A. 

\begin{table}[!h]
\centering
\caption{Comparison of the Individual and Aggregate Models ($S=500$)}
\label{tab:pjm}
\setlength{\tabcolsep}{3pt}
\begin{tabular}{lrrrrrrr}
\toprule
Model & $W^\star$ & $c^{\mathrm{F}}$ & $c_1$ & $c_2$ & $c_3$ & Binaries & Time (s)\\
\midrule
$\mathcal{P}_{\mathrm{ind}}$ & 298.28 & 200.0 & 98.70 & 8.26 & 4.06 & 4000 & 5801.5\\
$\mathcal{P}_{\mathrm{agg}}$ & 298.28 & 200.0 & 98.70 & 8.26 & 4.06 & 1500 & 1.56\\
\bottomrule
\end{tabular}
\end{table}

\vspace{-1em}
\begin{table}[!ht]
\centering
\caption{Marginal Upgrade Values from $\mathcal{P}_{\mathrm{LP}}$ ($S=2000$)}
\label{tab:duals}
\begin{tabular}{lrr}
\toprule
Facility & \shortstack{Line/generation\\capacity limits (MW)} & \shortstack{Marginal upgrade\\value $\lambda$}\\
\midrule
Line D--E & 240 & 1.92\\
Generation at D & 200 & 0.92\\
Generation at C & 520 & 0.62\\
Generation at A & 210 & 0.21\\
\bottomrule
\end{tabular}
\end{table}



\begin{figure}[!h]
\centering
\includegraphics[width=0.85\columnwidth]{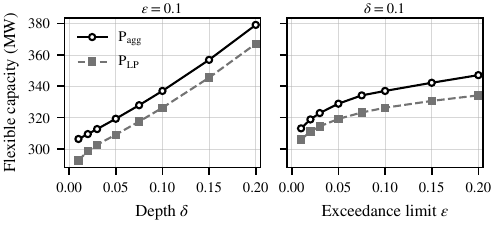}
\vspace{-6pt}
\caption{Effect of curtailment depth $\delta$ (left, $\epsilon=0.1$) and exceedance limit $\epsilon$ (right, $\delta=0.1$) on single-tier flexible capacity ($S=2000$, $c^{\mathrm{F}}=0$).}
\label{fig:delta-eps}
\vspace{-6pt}
\end{figure}

Next, we study the two QoS parameters separately to see how each affects flexible capacity. With one flexible tier and no firm capacity, we maximize the flexible award, and the flexible request limit is relaxed.  Fig.~\ref{fig:delta-eps} compares the aggregate and CVaR-based LP results. Both admit more capacity when either $\delta$ or $\epsilon$ increases. In the aggregate model, raising $\delta$ from 0.01 to 0.20 increases the award from 306.4 to 379.1~MW (24\%), while raising $\epsilon$ from 0.01 to 0.20 increases it by 11\%. The LP award is approximately 96.8\% of the aggregate award. 
\vspace{-1.7em}


\subsection{IEEE RTS 73-Bus System}

We also test the method on the PGLib-OPF IEEE RTS 73-bus system~\cite{pglib}, using 24-h  trajectories~\cite{grigg1999rts}.  We apply multiplicative Gaussian perturbations at the daily level ($\sigma=2\%$) and independently across bus-hours ($\sigma=3\%$), truncating both at $\pm 2\sigma$. Three data centers, at buses 103, 214, and 319, each request up to 400~MW of firm capacity and 200~MW per flexible tier, with the same QoS parameters and weights as in the PJM case. Global optimality is not proven for either $\mathcal{P}_{\mathrm{ind}}$ or $\mathcal{P}_{\mathrm{agg}}$ within two hours. We therefore report the results of $\mathcal{P}_{\mathrm{LP}}$, which is solved in 245~s.


\begin{table}[!ht]
\centering
\caption{RTS-73 Capacity Awards (MW) and Maximum In-Sample Depth-Exceedance Rates ($S=2000$)}
\label{tab:rts}
\setlength{\tabcolsep}{3pt}
\begin{tabular}{lrrrr}
\toprule
 & $c^{\mathrm{F}}$ & Tier 1 & Tier 2 & Tier 3\\
\midrule
Bus 103 & 400.0 & 0.0 & 0.0 & 55.8\\
Bus 214 & 400.0 & 0.0 & 0.0 & 128.8\\
Bus 319 & 400.0 & 144.9 & 154.0 & 200.0\\
Total & 1200.0 & 144.9 & 154.0 & 384.6\\
\midrule
Max. in-sample (\%) & -- & 0.15 & 0.70 & 3.75\\
\bottomrule
\end{tabular}
\end{table}

For each active product, we recover individual curtailments using~\eqref{eq:recovery} and compute the fraction of the 2000 trajectories in which its allowed depth is exceeded. Table~\ref{tab:rts} reports the maximum of these fractions across active products in each tier. All rates remain below their limits, consistent with the in-sample guarantee of Proposition~\ref{thm:cvar-guarantee}.
All three firm requests are awarded in full, and the total admitted capacity is 1883.5~MW. This is 45.0\% above the 1298.8~MW firm-only benchmark, computed on the same demand trajectories with relaxed firm-request upper bounds.


\section{Conclusion}\label{sec:conclusion}
This paper studies large-load interconnection with firm capacity and multiple flexible capacity tiers under a two-layer curtailment priority rule. We show that the individual and aggregate quantification models have the same feasible set of capacity decisions. This removes the priority binaries without changing the optimal value, thus supporting the efficient quantification of hosting capacity. The CVaR approximation further yields an LP that satisfies the sampled individual QoS limits and provides dual information for screening possible grid upgrades.



\nobalance
\appendices
\section{Proof of Proposition~\ref{prop:stack}}\label{app:proof-stack}
\begin{proof} 
Fix $z\in[0,C_i]$ and define $x_{im}$ by
\eqref{eq:tranche-recovery}.
Set $\beta_{i0}=1$, $\beta_{i,2K}=0$, and
$\beta_{im}\defeq\mathbf 1\{z\ge\Gamma_{im}\}$
for $0<m<2K$.
Here, $\mathbf 1\{\cdot\}$ denotes the indicator function.
Obtain $y_{ik}$ from~\eqref{eq:product-recovery-row}. Since the intervals $[\Gamma_{i,m-1},\Gamma_{im}]$ cover $[0,C_i]$,~\eqref{eq:tranche-recovery} allocates $z$ across the tranches in order, giving $0\le x_{im}\le d_{im}$ and $\sum_m x_{im}=z$. We verify that this allocation satisfies the priority constraints. If $\beta_{i,m-1}=0$, then $z<\Gamma_{i,m-1}$ and $x_{im}=0$; otherwise, $x_{im}\le d_{im}\le M$. Thus,~\eqref{eq:tranche-open} holds. Similarly, if $\beta_{im}=1$, then $z\ge\Gamma_{im}$ and $x_{im}=d_{im}$; otherwise, the lower bound in~\eqref{eq:tranche-full} is nonpositive. Since $\Gamma_{im}$ is nondecreasing in $m$, the binary variables satisfy~\eqref{eq:binary-chain} and~\eqref{eq:binary-endpoints}. Finally,~\eqref{eq:product-recovery-row} assigns each tranche to one product, so $\sum_k y_{ik}=\sum_m x_{im}=z$. Hence all constraints in~\eqref{eq:priority-representation} are satisfied.
Conversely, consider any feasible $(\mathbf x,\mathbf y,\boldsymbol\beta)$ with total curtailment $z$. By~\eqref{eq:binary-chain} and~\eqref{eq:binary-endpoints}, the binary variables form a nonincreasing sequence from $1$ to $0$. Let $m_0\defeq\min\{m:\beta_{im}=0\}$. For $m<m_0$, constraint~\eqref{eq:tranche-full} forces $x_{im}=d_{im}$, while for $m>m_0$, constraint~\eqref{eq:tranche-open} forces $x_{im}=0$. Thus, only tranche $m_0$ can be partially curtailed. Since $\sum_m x_{im}=z$, we must have
\[
x_{i,m_0}=z-\Gamma_{i,m_0-1},
\qquad
\Gamma_{i,m_0-1}\le z\le\Gamma_{i,m_0}.
\]
Therefore, all tranche curtailments are uniquely determined by $z$ and are given by~\eqref{eq:tranche-recovery}.

 Each product $k$ consists of tranches $K-k+1$ and $2K-k+1$, with sizes $a_{ik}$ and $b_{ik}$ and starting points $A_{ik}$ and $H_{ik}$, respectively. Applying~\eqref{eq:tranche-recovery} to these two tranches and summing their curtailments gives~\eqref{eq:recovery}.
\end{proof}

\section{Proof of Proposition~\ref{prop:event}}\label{app:proof-event}
\begin{proof}
For an active product, $b_{ik}=(1-\delta_k)c_{ik}>0$. By definition, $H_{ik}$ is the aggregate curtailment level at which the exceedance tranche of product $k$ starts. Moreover, from~\eqref{eq:threshold},
\[
H_{ik}-A_{ik}-a_{ik}
=\sum_{r<k}a_{ir}+\sum_{r>k}b_{ir}\ge0,
\]
so the within-depth tranche is fully curtailed before the exceedance tranche can be used.

Fix $z\in[0,C_i]$. If $z\le H_{ik}$, the exceedance tranche receives no curtailment, and~\eqref{eq:recovery} gives $f_{ik}(z;\mathbf c_i)\le a_{ik}=\delta_kc_{ik}$. If $z>H_{ik}$, the within-depth tranche is fully curtailed and the exceedance tranche receives positive curtailment, so $f_{ik}(z;\mathbf c_i)>\delta_kc_{ik}$. Hence, $f_{ik}(z;\mathbf c_i)>\delta_kc_{ik}$ if and only if $z>H_{ik}$.

By Proposition~\ref{prop:stack}, $y_{ikt}=f_{ik}(z_{it};\mathbf c_i)$ in every period. Therefore, product $k$ exceeds its allowed depth during the planning horizon if and only if the aggregate curtailment exceeds $H_{ik}$ in at least one period:
\[
\max_{t\in\T}y_{ikt}>\delta_kc_{ik}
\iff \exists\,t\in\T:\ z_{it}>H_{ik}
\iff Z_i>H_{ik}.
\]
This establishes~\eqref{eq:event}. Since the two events coincide in every realization, their probabilities are equal, which proves the equivalence of the two chance constraints.
\end{proof}

\section{Proof of Theorem~\ref{thm:equivalence}}\label{app:proof-equivalence}
\begin{proof}
\emph{From $\mathcal{P}_{\mathrm{ind}}$ to $\mathcal{P}_{\mathrm{agg}}$.} Take a feasible solution of $\mathcal{P}_{\mathrm{ind}}$ and set $z_{it}\defeq\sum_ky_{ikt}$. By~\eqref{eq:tranche-bounds} and~\eqref{eq:product-recovery-row}, $0\le y_{ikt}\le a_{ik}+b_{ik}=c_{ik}$, so~\eqref{eq:z-bounds} holds. Substituting~\eqref{eq:aggregate-def} into~\eqref{eq:individual-injection} gives~\eqref{eq:aggregate-injection} with the same nodal injections, while~\eqref{eq:balance}--\eqref{eq:ramp} remain unchanged. For each active product, Proposition~\ref{prop:event} shows that~\eqref{eq:individual-risk-row} implies~\eqref{eq:aggregate-risk}. For an inactive product $(i,k)$, suppose there is an active tier with index smaller than $k$ and let $h\defeq \max\{r<k:c_{ir}>0\}$. All tiers $r$ with $h<r\le k$ are inactive, so $H_{ik}=H_{ih}$ by~\eqref{eq:threshold}. Since $\epsilon_h\le\epsilon_k$ by~\eqref{eq:order}, we have $\Prob(Z_i>H_{ik})=\Prob(Z_i>H_{ih})\le\epsilon_h\le\epsilon_k$. If no such $h$ exists, then $c_{ir}=0$ for all $r\le k$, so $H_{ik}=C_i\ge Z_i$ and the exceedance event is empty.

\emph{From $\mathcal{P}_{\mathrm{agg}}$ to $\mathcal{P}_{\mathrm{ind}}$.} Take a feasible solution of $\mathcal{P}_{\mathrm{agg}}$. For each candidate bus, period, and realization,~\eqref{eq:z-bounds} gives $z_{it}\in[0,C_i]$. By Proposition~\ref{prop:stack}, set $y_{ikt}:=f_{ik}(z_{it};\mathbf c_i)$ and obtain $x_{imt}$ from~\eqref{eq:tranche-recovery}. Set $\beta_{i0t}=1$, $\beta_{i,2K,t}=0$, and $\beta_{imt}:=\mathbf 1\{z_{it}\ge\Gamma_{im}\}$ for $0<m<2K$. These variables satisfy~\eqref{eq:priority-representation} with $\sum_ky_{ikt}=z_{it}$. The recovered $\mathbf x$ and $\mathbf y$ are continuous functions of $z_{it}$, and the binary variables are indicator functions of $z_{it}$, so these are valid recourse decisions. Since $\sum_ky_{ikt}=z_{it}$,~\eqref{eq:individual-injection} gives the same nodal injections, and~\eqref{eq:balance}--\eqref{eq:ramp} remain satisfied. For each active product, Proposition~\ref{prop:event} turns~\eqref{eq:aggregate-risk} into~\eqref{eq:individual-risk-row}. For an inactive product, $y_{ikt}=0$, so its exceedance event is empty. The recovered solution is therefore feasible for $\mathcal{P}_{\mathrm{ind}}$.

Both mappings preserve $(\mathbf c^{\mathrm{F}},\mathbf c)$, so the two problems have the same feasible set of capacity decisions. Since their common objective $W$ depends only on these capacities, their optimal values are also equal, and an optimal solution of either problem is mapped to an optimal solution of the other.
\end{proof}

\section{Derivation of the CVaR-Based Linear Program}
\label{app:cvar-lp-derivation}
The loss of product $(i,k)$ in trajectory $s$ is
\begin{equation}
L_{ik}^s\defeq\max_{t\in\T}z_{it}^s-H_{ik}(\mathbf c_i).
\label{eq:app-scenario-loss}
\end{equation}
The minimization representation of CVaR~\cite{rockafellar2000} gives
\begin{equation}
\cvar_{1-\epsilon_k}(L_{ik})
=\min_{\tau_{ik}\in\R}
\left\{\tau_{ik}+\frac{1}{\epsilon_k}
\sum_{s\in\Sc}p_s\pos{L_{ik}^s-\tau_{ik}}\right\}.
\label{eq:app-cvar-representation}
\end{equation}
For a finite scenario set and $0<\epsilon_k<1$, this minimum is attained. Hence, $\cvar_{1-\epsilon_k}(L_{ik})\le0$ holds if and only if there exist $\tau_{ik}\in\R$ and auxiliary variables $\xi_{ik}^s$ such that
\begin{subequations}\label{eq:app-cvar-epigraph}
\begin{align}
&\tau_{ik}+\frac{1}{\epsilon_k}
\sum_{s\in\Sc}p_s\xi_{ik}^s\le0,
\label{eq:app-cvar-budget}\\
&\xi_{ik}^s\ge\pos{L_{ik}^s-\tau_{ik}},
\quad\forall s\in\Sc.
\label{eq:app-cvar-excess}
\end{align}
\end{subequations}
Indeed, a feasible CVaR constraint admits a minimizing $\tau_{ik}$ and the choice $\xi_{ik}^s\defeq\pos{L_{ik}^s-\tau_{ik}}$. Conversely, any variables satisfying~\eqref{eq:app-cvar-epigraph} provide an upper bound on~\eqref{eq:app-cvar-representation} that is at most zero.

Using~\eqref{eq:app-scenario-loss}, the excess loss can be written as
\begin{equation}
\begin{split}
\pos{L_{ik}^s-\tau_{ik}}
=\max\Big\{0,\max_{t\in\T}
\big\{z_{it}^s-H_{ik}(\mathbf c_i)-\tau_{ik}\big\}\Big\}.
\end{split}
\label{eq:app-max-excess}
\end{equation}
Therefore,~\eqref{eq:app-cvar-excess} is equivalent to
\begin{equation}
\begin{aligned}
\xi_{ik}^s&\ge0,\\
\xi_{ik}^s&\ge z_{it}^s-H_{ik}(\mathbf c_i)-\tau_{ik},
\quad\forall t\in\T.
\end{aligned}
\label{eq:app-linear-excess}
\end{equation}
Thus, one variable $\xi_{ik}^s$ per product and trajectory accounts for the maximum over all periods. Imposing~\eqref{eq:app-cvar-budget} and~\eqref{eq:app-linear-excess} for every $(i,k)$ and $s$ yields~\eqref{eq:epigraph1}--\eqref{eq:epigraph2} and $\xi_{ik}^s\ge0$.

Retaining the capacity bounds~\eqref{eq:bounds} and imposing the network constraints for each trajectory gives~\eqref{eq:lp-injection}--\eqref{eq:lp-ramp}. Since the objective is linear and $\mathbf C(\mathbf c)$ and $H_{ik}(\mathbf c_i)$ are affine in the capacity decisions, the resulting formulation is an LP. Introducing $\boldsymbol\tau$ and $\boldsymbol\xi$ preserves the objective and feasible decisions in the original variables, so $\mathcal{P}_{\mathrm{LP}}$ is equivalent to the finite-scenario CVaR formulation.

\section{Proof of Proposition~\ref{thm:cvar-guarantee}}\label{app:proof-cvar}
\begin{proof}
Fix $(i,k)$, write $\epsilon\defeq \epsilon_k$ and $\tau\defeq \tau_{ik}$, and let $L^s\defeq \max_{t}\{z_{it}^s-H_{ik}(\mathbf c_i)\}$. By~\eqref{eq:epigraph1} and~\eqref{eq:excess-domains}, $\xi_{ik}^s\ge\pos{L^s-\tau}$. Thus,~\eqref{eq:epigraph2} gives
\[
\tau+\frac1\epsilon\sum_{s\in\Sc}p_s\pos{L^s-\tau}\le0,
\]
which implies $\tau\le0$. If $\tau=0$, then $L^s\le0$ for every $s$ because $p_s>0$. If $\tau<0$, then $\pos{L^s-\tau}\ge(-\tau)\mathbf 1\{L^s>0\}$, and therefore
\begin{align*}
\Prob\big(Z_i>H_{ik}(\mathbf c_i)\big)&=\sum_{s\in\Sc}p_s\mathbf 1\{L^s>0\}\\
&\le\frac{1}{-\tau}\sum_{s\in\Sc}p_s\pos{L^s-\tau}\le\epsilon.
\end{align*}
In both cases,~\eqref{eq:aggregate-risk} holds. The remaining constraints of $\mathcal{P}_{\mathrm{agg}}$ are also satisfied by $\mathcal{P}_{\mathrm{LP}}$, so the solution is feasible for $\mathcal{P}_{\mathrm{agg}}$. Theorem~\ref{thm:equivalence} then gives feasibility for $\mathcal{P}_{\mathrm{ind}}$ and the product-level reliability guarantee.
\end{proof}

\end{document}